\documentclass[twocolumn]{aastex631}
\graphicspath{{./fig/}}

\usepackage{color}

\usepackage{bm}

\usepackage{ulem}
\usepackage{xcolor}

\begin{document}

\title{Early water delivery to terrestrial planet regions during the stages of Jupiter's formation and migration in the Grand Tack model}

\author[0000-0002-8300-7990]{Masahiro Ogihara}
\affiliation{Tsung-Dao Lee Institute, Shanghai Jiao Tong University, 520 Shengrong Road, Shanghai 201210, China}
\affiliation{Earth-Life Science Institute, Tokyo Institute of Technology, Meguro, Tokyo 152-8550, Japan}

\author[0000-0001-6702-0872]{Hidenori Genda}
\affiliation{Earth-Life Science Institute, Tokyo Institute of Technology, Meguro, Tokyo 152-8550, Japan}

\author[0000-0002-3544-3632]{Yasuhito Sekine}
\affiliation{Earth-Life Science Institute, Tokyo Institute of Technology, Meguro, Tokyo 152-8550, Japan}
\affiliation{Institute of Nature and Environmental Technology, Kanazawa University, Ishikawa, 920-1192, Japan}



\begin{abstract}
The formation and subsequent migration of gas giants could significantly affect the material mixing in the Solar System. In this study, we use \textit{N}-body simulations to investigate how much water is transported into the region of the terrestrial planet formation during the growth and migration phases of Jupiter in the Grand Tack model. We found that Jupiter's growth was accompanied by significant mass transport, and that a substantial amount of water (about 10 times Earth's ocean mass for the initial planetesimal distribution based on the minimum-mass solar nebula) was transported into the terrestrial planet region. The total amount delivered increased further during Jupiter's migration phase (totaling about 10--40 times Earth's ocean mass), which was less dependent on simulation parameters. In addition, at these stages, terrestrial planets were not fully grown. Therefore, water supplied during these early stages could interact with metallic iron during the core formation of protoplanets and/or growing Earth. Since hydrogen in water molecules can dissolve into their cores, this could explain the density deficit observed in the current Earth core. Notably, Jupiter could play an important role as a ``barrier'' in explaining the dichotomy of the isotopic compositions between noncarbonaceous (NC) and carbonaceous (CC) meteorites. This study's results show that Jupiter's growth necessitates some mixing of NC and CC materials.
\end{abstract}



\section{Introduction}\label{sec:intro}
Material mixing is thought to have occurred during the Solar System formation, with materials migrating from one location to another. For example, materials that formed Earth are considered mostly noncarbonaceous (NC) \citep[e.g.,][]{2017Natur.541..521D}. However, some amount of carbonaceous (CC) material formed at greater distances from Earth may have been accreted to Earth \citep{2018Natur.555..507S}. The Earth's ocean may have originated from carbonaceous chondrites \citep[e.g.,][]{2012E&PSL.313...56M}.
Studies have been conducted to elucidate material mixing during and after the Solar System formation \citep[e.g.,][]{2000M&PS...35.1309M,2014Icar..239...74O}. 
In addition, Solar System explorations such as Stardust, Dawn, and Hayabusa2 have provided information on material distribution during the Solar System formation \citep{2008Sci...321.1664N,2015Natur.528..241D,2022NatAs...6..214Y}. Therefore, to decipher the observational and sample analysis data obtained by such explorations, insights into material mixing during Solar System formation are highly needed.

Material mixing in the Solar System is expected from the theoretical studies' perspectives. Although planetesimals 1--1000\,km in size do not migrate much due to gas drag \citep{1976PThPh..56.1756A}, gas giant planets can significantly affect planetesimal orbits.
Giant planets, when they grow by envelope accretion, can greatly alter the surrounding planetesimal orbits \citep{2017Icar..297..134R,2020PSJ.....1...45C}. In addition, planetesimal orbits are affected when giant planets undergo type II migration due to their interaction with the gas disk \citep{1999Icar..139..350T,2005A&A...441..791F,2006Sci...313.1413R,2011Natur.475..206W,2015ApJ...813...72C}.

The growth and migration of giant planets, such as Jupiter, depends on the Solar System formation model \citep[e.g.,][]{2022ASSL..466....3R}. One of the models is the Grand Tack model \citep{2011Natur.475..206W}, which postulates that Jupiter and Saturn migrate inward due to the disk-planet interaction and then change the direction to migrate outward after Jupiter's orbit approaches an orbit with a semimajor axis of about 1.5\,au.
Although the Grand Tack model suffers some potential problems\footnote{One caveat is that the model assumes that little or no gas accretion to giant planets occurs during their outward migration \citep{2012ApJ...757...50D}. If growth during migration is considered, the gas planets would exceed the current Jupiter or Saturn masses \citep{2020ApJ...891..143T}.  Notably, fine tuning may be required so that Jupiter and Saturn are captured in a 3:2 mean-motion resonance \citep{2014ApJ...795L..11P,2020MNRAS.492.6007C}.}, it remains a promising model because it explains the origin of several features of the Solar System, such as the features in the asteroid belt (e.g., mass depletion, high eccentricity, zoning) and small Mars.

The effect of the growth and migration of giant planets on planetesimal orbit has been studied from several viewpoints in recent years.
\citet{2017Icar..297..134R} investigated the effect of planetesimal scattering during Jupiter's growth, focusing mainly on the effect on the asteroid belt. They only considered giant planets growing at orbital radii of 5.4\,au or greater, and also did not investigate planetesimal delivery during Jupiter's outward migration.
\citet{2011Natur.475..206W} examined planetesimal transport during the inward and outward migration of giant planets, but did not consider water delivery by planetesimals during the growth phase of giant planets.
Note that \citet{2014Icar..239...74O} performed long-term (150\,Myr) \textit{N}-body simulations of planetesimals with perihelion distance of less than 2\,au after the dissipation of the protoplanetary disk using the final state of the simulation of \citet{2011Natur.475..206W}. They examined the accretion process of these planetesimals to terrestrial planets.
\citet{2015ApJ...813...72C} studied the growth of planetesimals during the migration of Jupiter in accordance with the Grand Tack model, focusing on planetesimals only in the inner orbit of Jupiter in the initial state.
\citet{2020PSJ.....1...45C} considered the growth and migration of Jupiter in the Grand Tack model and examined collisions between planetesimals and the delivery of planetesimals into the asteroid belt.
An investigation of the extent to which water was transported through Jupiter's growth and migration in the Grand Tack model is lacking, as is a discussion of the important implications associated with it.
In this study, we follow the growth and migration phases of Jupiter following the Grand Tack model and perform \textit{N}-body simulation to examine how much water and other materials are delivered to the Earth's orbit during these phases.
Note also that although some simulations calculate up to 10\,Myr, we do not follow the long-term ($\sim 100{\rm \,Myr}$) accretion of delivered planetesimals on terrestrial planets as investigated in \citet{2014Icar..239...74O}.

This paper is structured as follows.
In Section~\ref{sec:methods}, we describe our model of \textit{N}-body simulation.
In Section~\ref{sec:results}, we present the simulation results. We divide the simulation into two phases - the growth and migration phases-and examine the material mixing in each phase. 
Section~\ref{sec:discussion} discusses some important points inferred from our results. First, water transport to the Earth's core is discussed. One possible explanation for the density deficit in the Earth's core \citep{1952JGR....57..227B} is that water was transported during the Earth formation stage. It is unlikely that water transport to the core occurred by the conventional water delivery mechanism - the late accretion after the last giant impact.
This study's results suggest that early water delivery could transport water to the Earth's core.
Second, the NC-CC dichotomy is discussed. Isotope measurements of meteorites suggest that the Solar System was spatially divided into two planetesimal-forming regions for about several million years (Myr) \citep{2011E&PSL.311...93W}. This isolation can be attributed to Jupiter acting as a ``barrier'' to prevent mixing of NC and CC material \citep{2017PNAS..114.6712K}. 
Our results indicate that Jupiter's growth mixes NC and CC materials to some extent.
Additionally, although this study focuses primarily on the influence of Jupiter on material mixing, we discuss simulation results involving a growing Saturn near Jupiter.
Finally, we also discuss the relationship to previous studies and how this study can change the previous understanding.
In Section~\ref{sec:conclusions}, we summarize our findings.

\section{Numerical methods}\label{sec:methods}

We numerically calculated the transport of planetesimals using \textit{N}-body simulation. The simulation was divided into two phases: Jupiter's growth phase (up to $t=1\,{\rm Myr}$) and Jupiter's migration phase (or the Grand Tack phase).

As initial conditions, we considered three types of objects.
A Jupiter's core of mass $M=10\,M_\oplus$ was placed at a semimajor axis of $a=3.5\,{\rm au}$ according to the Grand Tack model. Forty planetary embryos of $0.05\,M_\oplus$ were placed in the terrestrial planet formation region ($a=1-1.3\,{\rm au}$)\footnote{Previous \textit{N}-body simulations based on the Grand Tack model assumed that the inner boundary of the initial distribution of embryos is at $a=0.7 {\rm \,au}$ \citep[e.g.,][]{2014RSPTA.37230174J}. Although recent studies have suggested several mechanisms by which the inner boundary could be created at 0.7\,au \citep[e.g.,][]{2018A&A...612L...5O,2021ApJ...921L...5U}, it is not very clear how the 0.7\,au boundary is formed in the presence of type I migration. This needs to be further investigated. In our initial conditions, we consider the inner edge of the embryo distribution at $a=1 {\rm \,au}$. Embryos move inward by scattering and eccentricity damping, and the inner edge reaches at 0.7\,au when the Grand Tack migration starts at $t=1 {\rm \,Myr}$.}. Planetesimals were placed in the region of $a=1.3-6\,{\rm au}$, excluding the region near Jupiter's core about 2-3 Hill radii from the core as in \citet{2017Icar..297..134R}. The distribution of embryos and planetesimals was set to be consistent with the minimum-mass solar nebula: $\Sigma_{\rm d} = 10 [r/({\rm 1\,au})]^{-3/2}\,{\rm g\,cm^{-2}}$, where $r$ is the radial distance from the Sun. The initial eccentricities and inclinations follow the Rayleigh distribution with a dispersion of 0.01. The mass of planetesimals was assumed to be $0.005\,M_\oplus$, and in this case, the number of planetesimals was 1106. We also performed higher resolution simulations with $M=0.0025\,M_\oplus$ and $N=2212$. As performed in \cite{2017Icar..297..134R}, mutual interactions between planetesimals were not considered since the long-term simulation in this study focused on planetesimal transport.

Jupiter's growth was calculated according to the envelope Kelvin-Helmholtz contraction using Eqs.\,(41)--(42) of \citet[][]{2021A&A...650A.116M} \citep[see also][]{2010ApJ...714.1343H}, the gas inflow near the planet using Eqs.\,(15)--(16) of \citet{2020ApJ...892..124O} \citep[see also][]{2016ApJ...823...48T}, and the disk accretion using Eq.\,(19) of \citet[][]{2020ApJ...892..124O}. Jupiter stops gas accretion when it reaches Jupiter mass. In our fiducial setup, the Jupiter core grew to Jupiter mass in about 1\,Myr.
When Jupiter grows, growth can be limited by the disk accretion \citep[e.g.,][]{2018ApJ...867..127O} and becomes near linear growth.
To observe the effect of Jupiter's growth rate, we also considered the case where Jupiter grows quickly in about 0.1\,Myr  \citep[e.g.,][]{2014ApJ...786...21P} by changing the disk accretion rate.

For Jupiter's migration, migration in the Grand Tack model was mimicked based on \citet{2011Natur.475..206W} and \citet{2015ApJ...813...72C}. Jupiter moved inward on a timescale of 100\,kyr and moved outward after reaching $a=1.5\,{\rm au}$. During the outward migration, as stated in \citet{2011Natur.475..206W} and \citet{2015ApJ...813...72C}, the migration gradually slowed down by considering exponential decay. Jupiter arrived at its current orbit ($a=5.2\,{\rm au}$) at $t=1.5\,{\rm Myr}$.

For the gas disk model, although disk models that account for disk winds have recently gained attention \citep{2016A&A...596A..74S,2018A&A...615A..63O}, we used a simple two-component power-law disk model \citep[e.g.,][]{2015A&A...575A..28B,2019A&A...632A...7L}. In this model, the disk is divided into the inner region, where the temperature is determined by viscous heating (vis), and the outer region, where stellar irradiation (irr) determines the disk temperature. Each region has a dominant heating mechanism and switches from the viscous regime to the irradiation regime at a specific radius (see Eq.\,(14) of \citet{2019A&A...632A...7L} for location). The gas surface density in each region is expressed by
\begin{eqnarray}
\Sigma_{\rm g, vis} = 1320 \left(\frac{\dot{M}_{\rm g}}{10^{-9}\,M_\odot\,{\rm yr}^{-1}}\right)^{1/2} \left(\frac{\alpha}{10^{-4}}\right)^{-3/4} \nonumber \\
\times \left(\frac{r}{{\rm 1\,au}}\right)^{-3/8}\,{\rm g\,cm^{-2}},\\
\Sigma_{\rm g, irr} = 2500 \left(\frac{\dot{M}_{\rm g}}{10^{-9}\,M_\odot\,{\rm yr}^{-1}}\right) \left(\frac{\alpha}{10^{-4}}\right)^{-1} \nonumber \\
\times \left(\frac{r}{{\rm 1\,au}}\right)^{-15/14}\,{\rm g\,cm^{-2}},
\end{eqnarray}
where we assume $\dot{M}_{\rm g} = 10^{-9} \exp[(t-{\rm 1\,Myr})/({\rm 0.5\,Myr})]\,M_\odot\,{\rm yr}^{-1}$ for the fiducial case and $\alpha = 10^{-4}$.
Jupiter opens a density gap in its vicinity, which was also considered in this study by using a model in which the depth and width of the gap is determined by giant planet mass and disk properties (e.g., viscosity) \citep{2017PASJ...69...97K,2019MNRAS.487.4510S}. 

As aerodynamical gas drag damps the eccentricity and inclination of planetesimals, we calculated the drag force by the following formula \citep{1976PThPh..56.1756A}.
\begin{equation}
\bm{F}_{\rm aero} = - \frac{C_{\rm D} \pi R^2 \rho_{\rm g} v_{\rm rel}}{2 M}  \bm{v}_{\rm rel},
\end{equation}
where $\rho_{\rm g}$ is the gas density, $v_{\rm rel}$ is the relative velocity between the planetesimal and gas, and $R$ is the radius of planetesimals and is varied between 5--500\,km to determine the effect of gas damping strength.
The drag coefficient $C_{\rm D}$ is calculated according to the Mach number, the Knudsen number, and the Reynolds number \citep[e.g.,][]{1976PThPh..56.1756A,2007Icar..191..413B}.

The focus of this paper is on water delivery. The amount of water transport was investigated in the post-process analysis. Therefore, the distribution of the water mass fraction of each planetesimal can be determined after performing \textit{N}-body simulations. The typical water mass fraction $\eta_{\rm w}$ used in this study followed previous \textit{N}-body simulations \citep[e.g.,][]{2004Icar..168....1R,2014ApJ...786...33Q,2020A&A...634A..76B} and arguments about meteorites \citep[e.g.,][]{2000orem.book..413A,2018ApJS..238...11D}, where planetesimals initially in $a < 2 \,{\rm au}$ had 0.001\% (enstatite chondrites), those in $2 \leq a < 3.5 \,{\rm au}$ had 0.1\% (ordinary chondrites), and those in $a \geq 3.5\,{\rm au}$ had 5\% (carbonaceous chondrites). This water distribution was based on the formation region of each chondrite.
Embryos are assumed to initially contain no water.
Note that the initial amount of water present in the terrestrial planet region ($a < 2 {\rm \,au}$) is 0.06 times the Earth ocean mass.

For the numerical integration\footnote{Simulation data used in this paper are available from the corresponding author upon reasonable request.}, we use the fourth-order Hermite scheme \citep[e.g.,][]{1992PASJ...44..141M,2004PASJ...56..861K} with the hierarchical timestep \citep{1991ApJ...369..200M}. We consider collisions between planets except for collisions between planetesimals. For all collisions, perfect merging is assumed.
Note that water can evaporate and dehydrate due to collisions between planetesimals. According to \citet{2020PSJ.....1...45C}, it is estimated that less than 10 percent of water-bearing planetesimals would experience collisions between planetesimals. Although many collisions exceed the onset of vaporization for ice \citep{2020PSJ.....1...45C}, collisions that cause dehydration of hydrous minerals are rare \citep{2019Icar..328...58W}.
Therefore, it is likely that such water loss would not have a large effect on the amount of water delivery investigated in this study. It would be interesting to examine the effect of water evaporation during collisions between planetesimals at smaller planetesimal sizes not studied in \citet{2020PSJ.....1...45C} in a future publication.

\section{Results}\label{sec:results}
\subsection{Growth phase}\label{sec:growth}

\begin{figure}[th]
\plotone{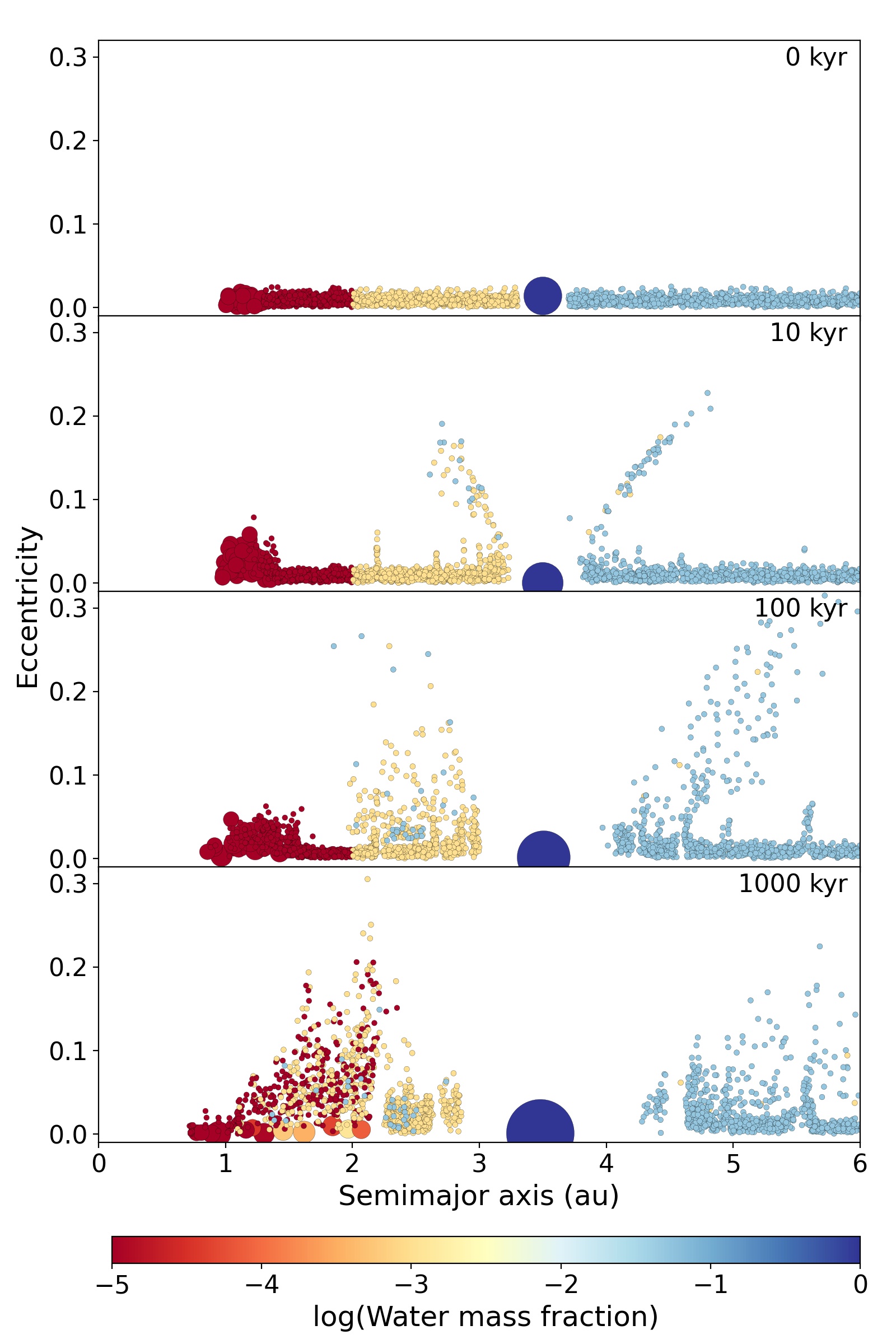}
\caption{Snapshots of the system during the growth phase. Jupiter grows at 3.5 au on a timescale of 1\,Myr and the size of the planetesimal is assumed to be 50\,km. The colors represent the water mass fraction of objects other than Jupiter. The water content of the initial planetesimals is described in Section~\ref{sec:methods}.
Notably, as Jupiter grows, the surrounding planetesimals are scattered, which contributes to the material mixing.}
\label{fig:snap_fid}
\end{figure}

We first consider the simulation results in the growth phase of Jupiter for our fiducial case, where Jupiter's growth is completed in about 1\,Myr and the size of planetesimals is 50\,km. Figure~\ref{fig:snap_fid} shows snapshots of a typical simulation up to 1\,Myr. As Jupiter grows, the planetesimals around Jupiter are scattered and their eccentricities are excited. Since the Jacobi energy is approximately conserved during the scattering, the planetesimals lie on the line of constant Jacobi energy inside and outside Jupiter ($t=10\,{\rm kyr}$).
The planetesimals with increased eccentricity are also subject to the aerodynamical drag force that damps their eccentricities. Therefore, the eccentricity of the planetesimal is determined by the balance of excitation by scattering and damping by gas drag. 
Notably, water-rich planetesimals (light blue) initially in Jupiter's outer orbits are scattered into the inner orbit.

\begin{figure}[th]
\plotone{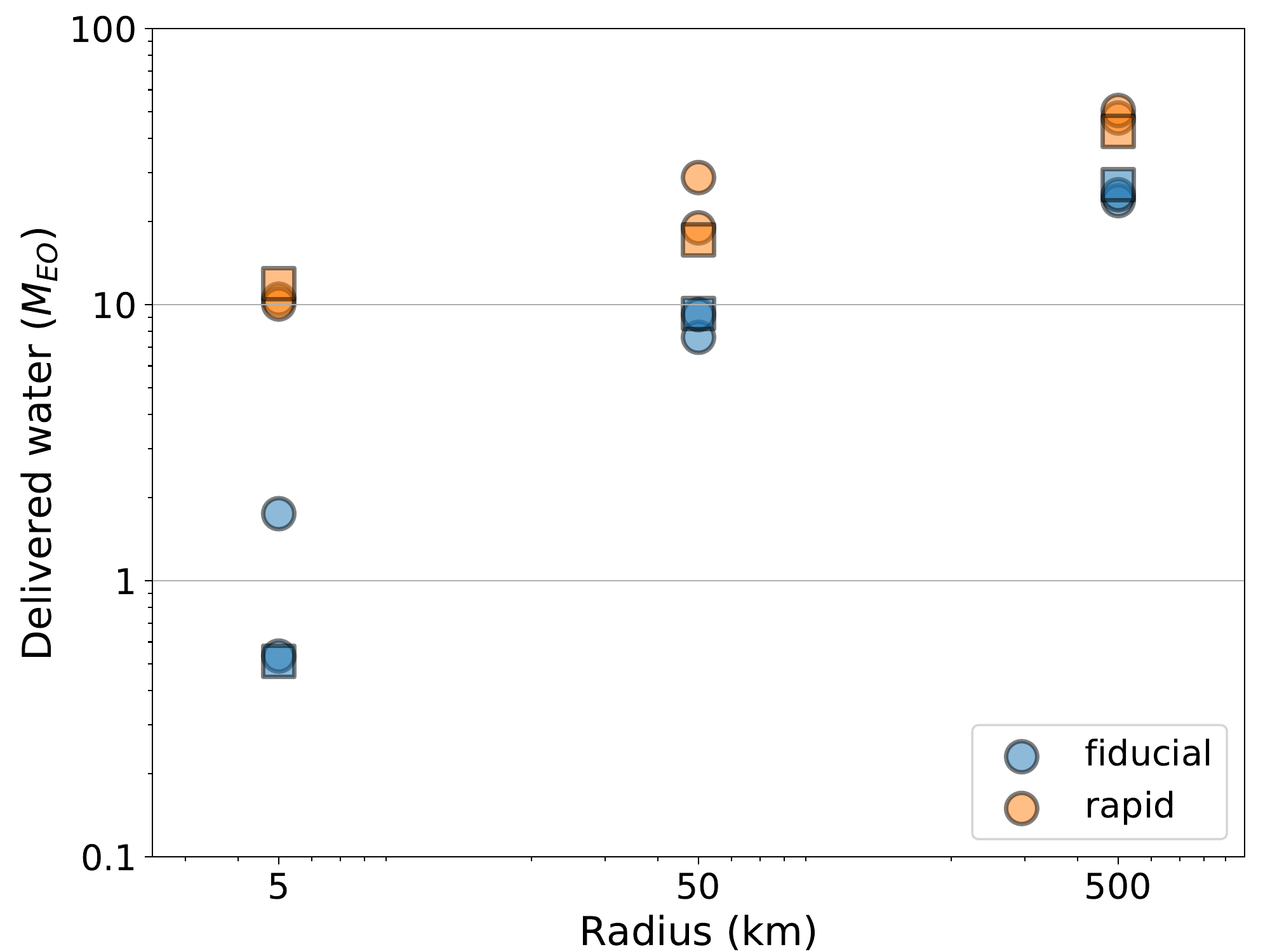}
\caption{The amount of water delivered into the region of $a<2\,{\rm au}$ during the growth phase of Jupiter in each simulation. The initial water mass fraction of the planetesimal is the value described in Section~\ref{sec:methods}. Three runs were performed in each setting. Squares represent results for higher resolution simulations.}
\label{fig:r_m}
\end{figure}

Figure~\ref{fig:r_m} shows how much water is delivered to the terrestrial planet-forming region\footnote{In this study, for simplicity, we consider planetesimals with $a < 2\,{\rm au}$ delivered to the terrestrial planetary region. We confirm that the conclusions of this paper do not change when the delivery is judged using the perihelion distance of planetesimals (perihelion distance of 2\,au or 1.5\,au). Longer ($\sim 100\,{\rm Myr}$) \textit{N}-body simulations are needed to determine whether they actually accrete on Earth after the delivery.} ($a<2\,{\rm au}$) until $t = 1\,{\rm Myr}$.
In the fiducial case with $R=50\,{\rm km}$ for planetesimals, we noticed that about 10\,$M_{\mathrm{EO}}$ were transported into that region, where $M_{\mathrm{EO}}$ is one Earth ocean mass corresponding to 0.00023 $M_\oplus$.
Thus, a significant amount of water-bearing planetesimals can be delivered to the terrestrial planet region during Jupiter's growth phase.

\begin{figure*}[th]
\plotone{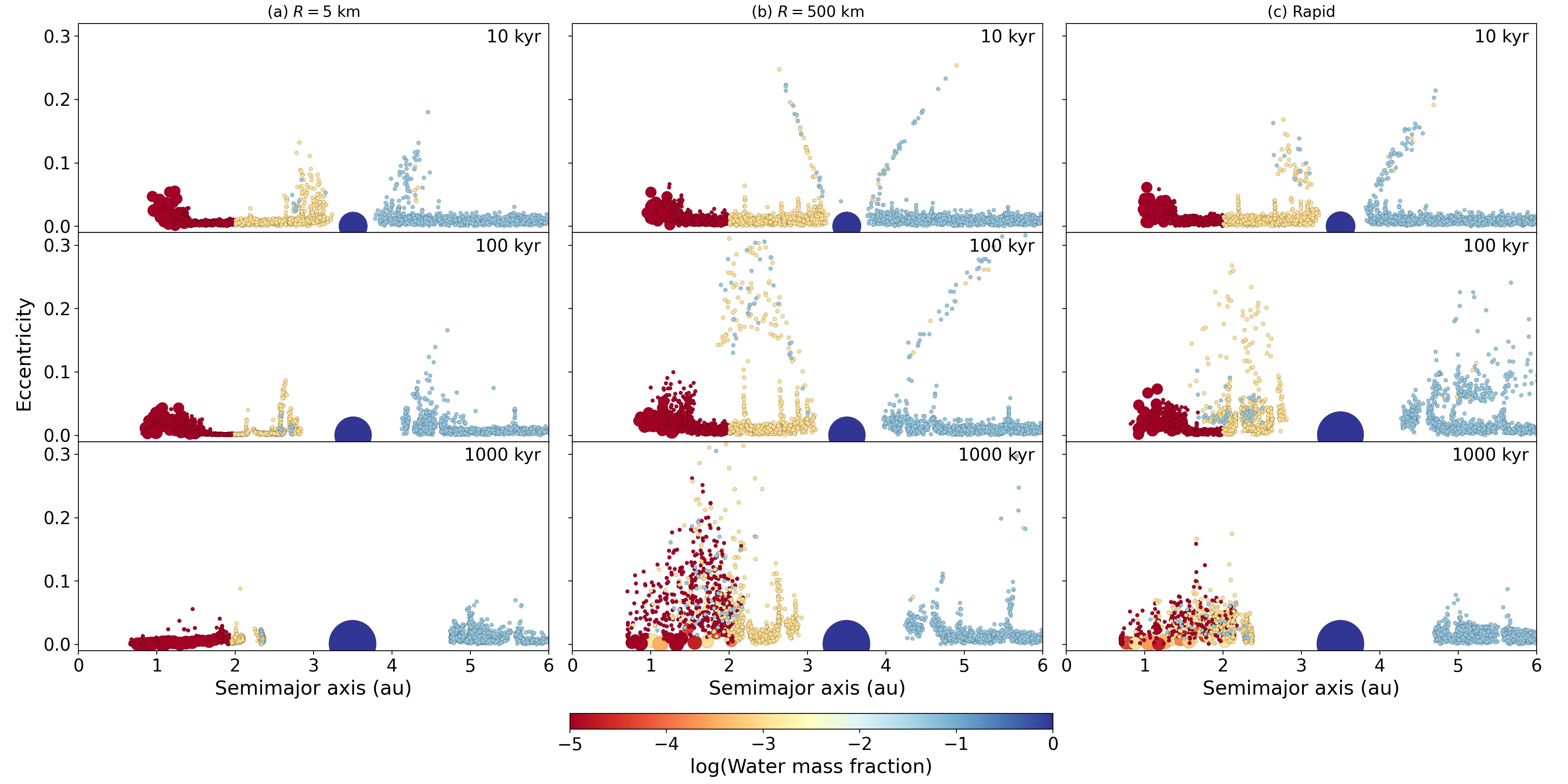}
\caption{Same as Figure~\ref{fig:snap_fid} but for different parameters.
(\textit{a}) The planetesimal is 5\,km in size and the eccentricity damping is strong.
(\textit{b}) The planetesimal is 500\,km in size and the eccentricity damping is weak.
(\textit{c}) Jupiter is in rapid growth with envelope accretion on a timescale of about 0.1\,Myr. The planetesimal is 50\,km in size.}
\label{fig:snap}
\end{figure*}

Furthermore, we show the simulation results with different parameters. Figure~\ref{fig:snap} (panels a and b) indicates the simulation results for different sizes of planetesimals.
Regarding the dependence on the planetesimal size, the transport efficiency of planetesimals was found to depend on the planetesimal size. When the planetesimal size was small and the eccentricity damping was strong, the damping had a stronger effect than the excitation by Jupiter's scattering. Consequently, scattering was less likely to cause migration in the semimajor axis, and the planetesimal transport was reduced. 
Conversely, when the planetesimal size was larger and the eccentricity-damping was weaker, scattering by Jupiter was more efficient and planetesimal transport was more pronounced.
Previous studies have also reported the dependence of planetesimal excitation on planetesimal size \citep[e.g.,][]{2017Icar..297..134R}.
Figure~\ref{fig:r_m} compares the amount of water delivery for different parameters. The amount of water transported to the terrestrial planet region is about 1\,$M_{\mathrm{EO}}$ for planetesimals with $R=5\,{\rm km}$, whereas it is about 20--30\,$M_{\mathrm{EO}}$ for planetesimals with $R=500\,{\rm km}$. Thus, the amount of water delivery during Jupiter's growth phase clearly depended on the planetesimal size (i.e., eccentricity damping).

When Jupiter grows faster (see Figure~\ref{fig:snap}(c)), the eccentricity of planetesimals scattered by Jupiter exceeds that seen in Figure~\ref{fig:snap_fid}. This makes the delivery of planetesimals more efficient.
The amount of delivered water in Figure~\ref{fig:r_m} confirmed that for rapid growth the water delivery was about 2--10 times higher.
In addition, a similar dependence on the planetesimal size was seen for rapid growth.

\subsection{Migration phase}
\begin{figure*}[th]
\epsscale{0.7}
\plotone{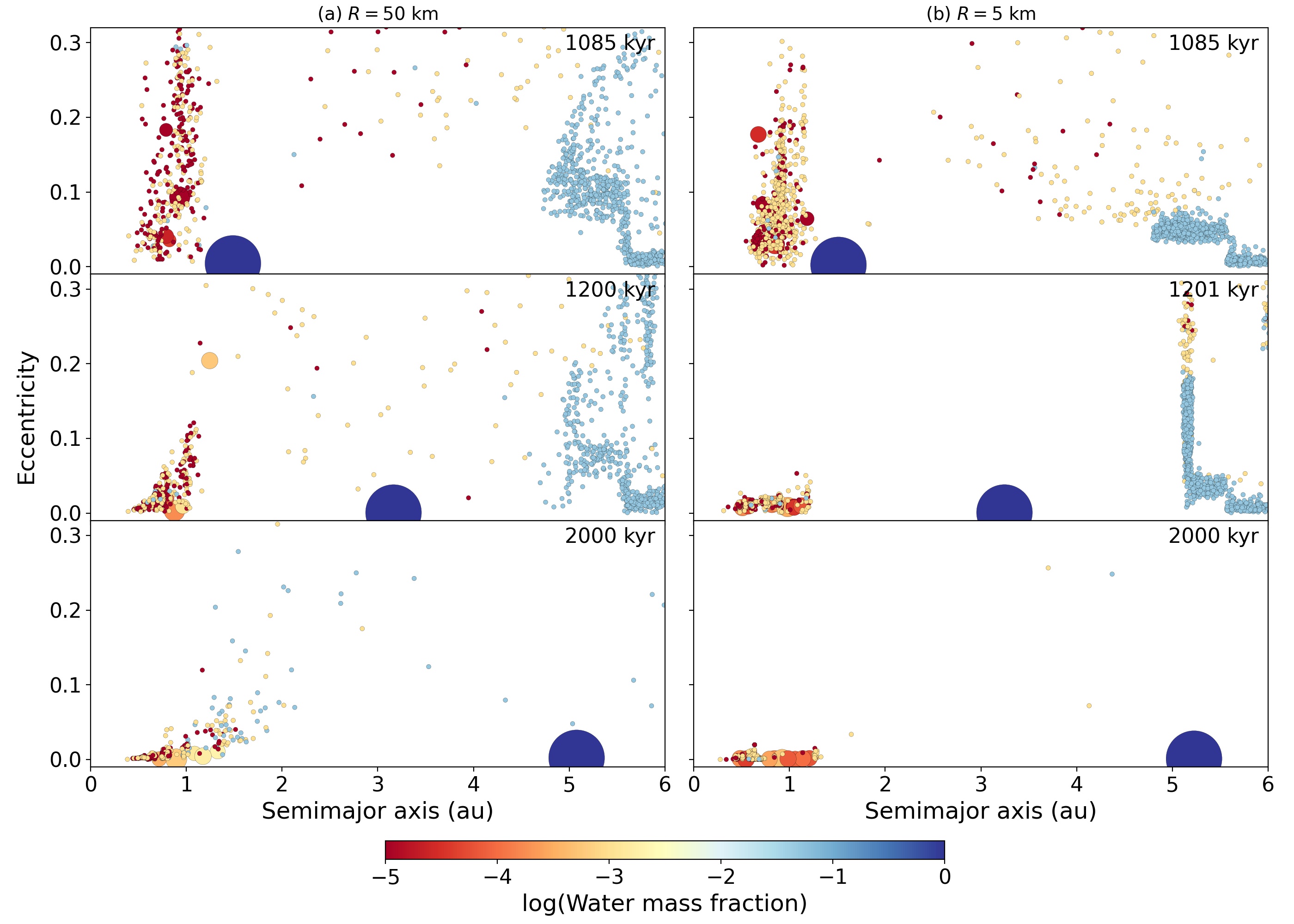}
\caption{Snapshots of the system during the migration phase. Jupiter moves inward and outward according to the Grand Tack model.
(\textit{a}) Our fiducial setting with planetesimal of 50\,km in size.
(\textit{b}) The planetesimal is 5\,km in size.
}
\label{fig:snap_mig}
\end{figure*}

We consider how mass transport occurs as Jupiter moves according to the Grand Tack model. This stage corresponds to the simulation by \citet{2011Natur.475..206W}. Notably, while \citet{2011Natur.475..206W} consider a spatially uniform distribution of planetesimals for initial conditions, the planetesimal distribution after Jupiter's growth significantly differed in our simulations. We used the final distribution of planetesimals obtained in Jupiter's growth phase as the initial distribution in Jupiter's migration phase.

Figure~\ref{fig:snap_mig}(a) shows a typical result of the migration phase for the fiducial case; a continuation case with the fiducial case in the growth phase. Most planetesimals transported from Jupiter's outer orbit to the inner orbit $(a<3.5\,{\rm au})$ during the Jupiter growth phase were swept into the more inner orbit during the Jupiter's inward migration ($t=1085\,{\rm kyr}$).
This results in additional mass transport to the inner region. At this stage, additional scattering of planetesimals from Jupiter's outer orbit into the inner orbit was limited. The main process is the further transport of planetesimals that are already in Jupiter's inner orbit.
Jupiter then changes its migration direction to outward and reaches its current orbit at $t=1500\,{\rm kyr}$. Finally, we observed that a significant amount of planetesimals in Jupiter's outer orbit were delivered to the terrestrial planet region.

Figure~\ref{fig:snap_mig}(b) shows an example of the migration phase when the planetesimal radius is small and eccentricity damping is strong. In addition, during Jupiter's inward migration, planetesimals at $a < 3.5\,{\rm au}$ were swept up to $a<1.5\,{\rm au}$.

\begin{figure}[th]
\plotone{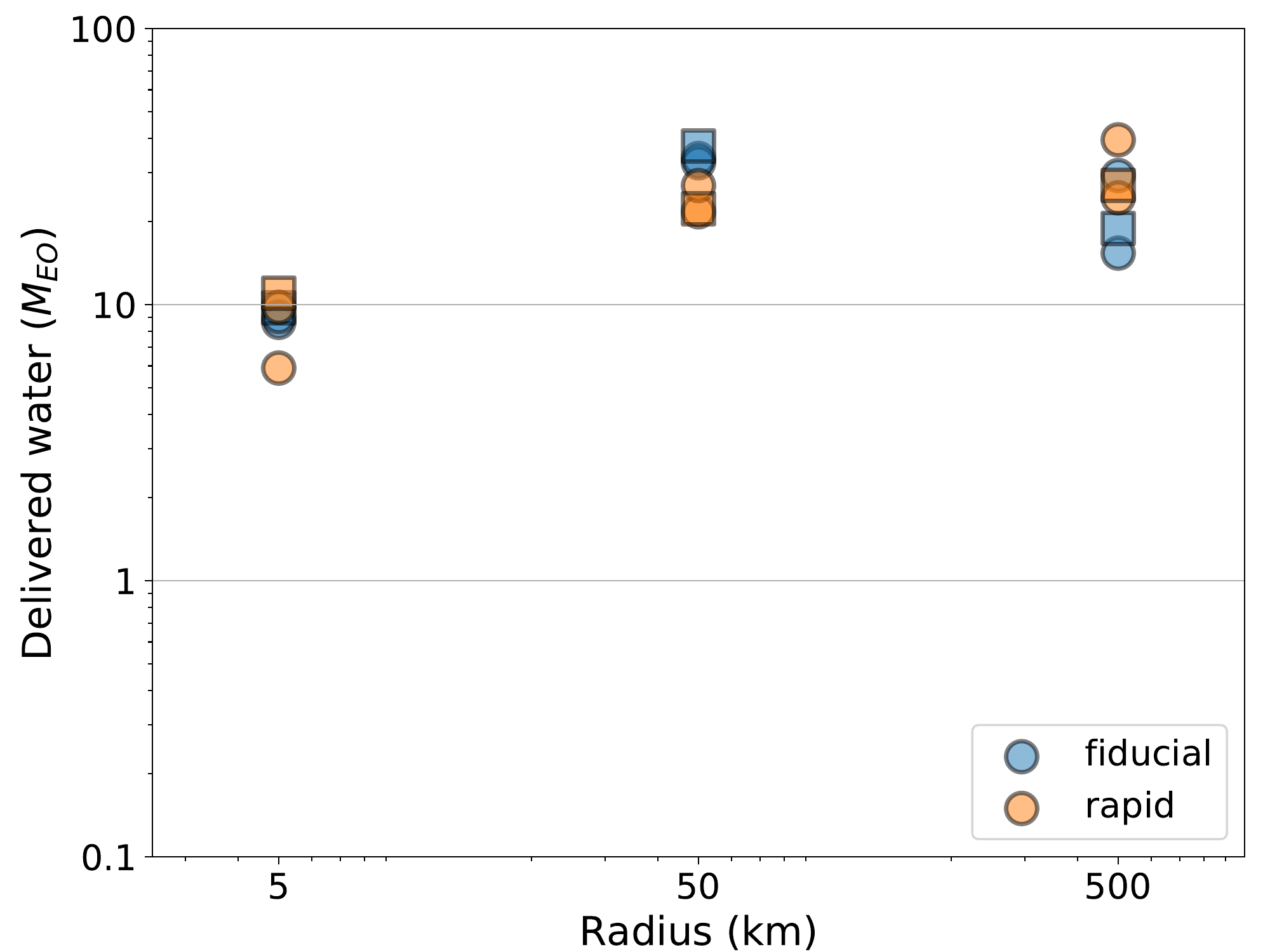}
\caption{Same as Figure~\ref{fig:r_m}, but the total amount of water delivered into the region of $a<2\,{\rm au}$ during the growth and migration phase of Jupiter. 
}
\label{fig:r_m_mig}
\end{figure}

We then considered the estimates of the amount of water delivered. Figure~\ref{fig:r_m_mig} shows the total amount of water transported to the terrestrial planet region through Jupiter's growth and migration phases. In the fiducial case with $50\,{\rm km}$-size planetesimals, we noticed that a large amount of water (about 20\,$M_{\mathrm{EO}}$) was delivered during the Grand Tack migration phase, which is on the same order of magnitude as the amount transported during Jupiter's growth phase (about 10\,$M_{\mathrm{EO}}$).
For $5\,{\rm km}$-size planetesimals, a large amount of water (about 9 $M_{\mathrm{EO}}$) was also delivered during the migration phase, which significantly exceeded the amount transported during the growth phase (about 1\,$M_{\mathrm{EO}}$). In total, the amount of water delivered was less than that for $R=50\,{\rm km}$.
Conversely, for larger planetesimals ($R=500\,{\rm km}$), little additional water was observed to be delivered during the migration phase. However, the number of planetesimals transported to $a < 2\,{\rm au}$ during the growth phase was somewhat reduced due to scattering during the Jupiter's migration. Consequently, the final water delivery between the cases $R=50\,{\rm km}$ and $R=500\,{\rm km}$ did not differ significantly.

In Figure~\ref{fig:r_m_mig}, the results when Jupiter was growing fast during the growth phase was considered. The basic trend is the same as when Jupiter grows more slowly. There is additional transport in the case of $R=5\,{\rm km}$, where little water was delivered during the growth phase. There is little additional water delivery in the case of $R=50/500\,{\rm km}$, where there was significant transport during the growth phase.
Notably, as in the fiducial case, the final transport between the cases of $R=50\,{\rm km}$ and $500\,{\rm km}$ did not differ.
In addition, the final water transport did not depend on Jupiter's growth timescale.
Therefore, the amount of water delivered did not depend much on the simulation parameters (about 10--40 $M_{\rm EO}$ in total) and can be divided into two categories: cases with and without strong eccentricity damping.

\subsection{Fitting formula for mass delivery}
\begin{figure}[th]
\begin{center}
\includegraphics[width=0.49 \textwidth]{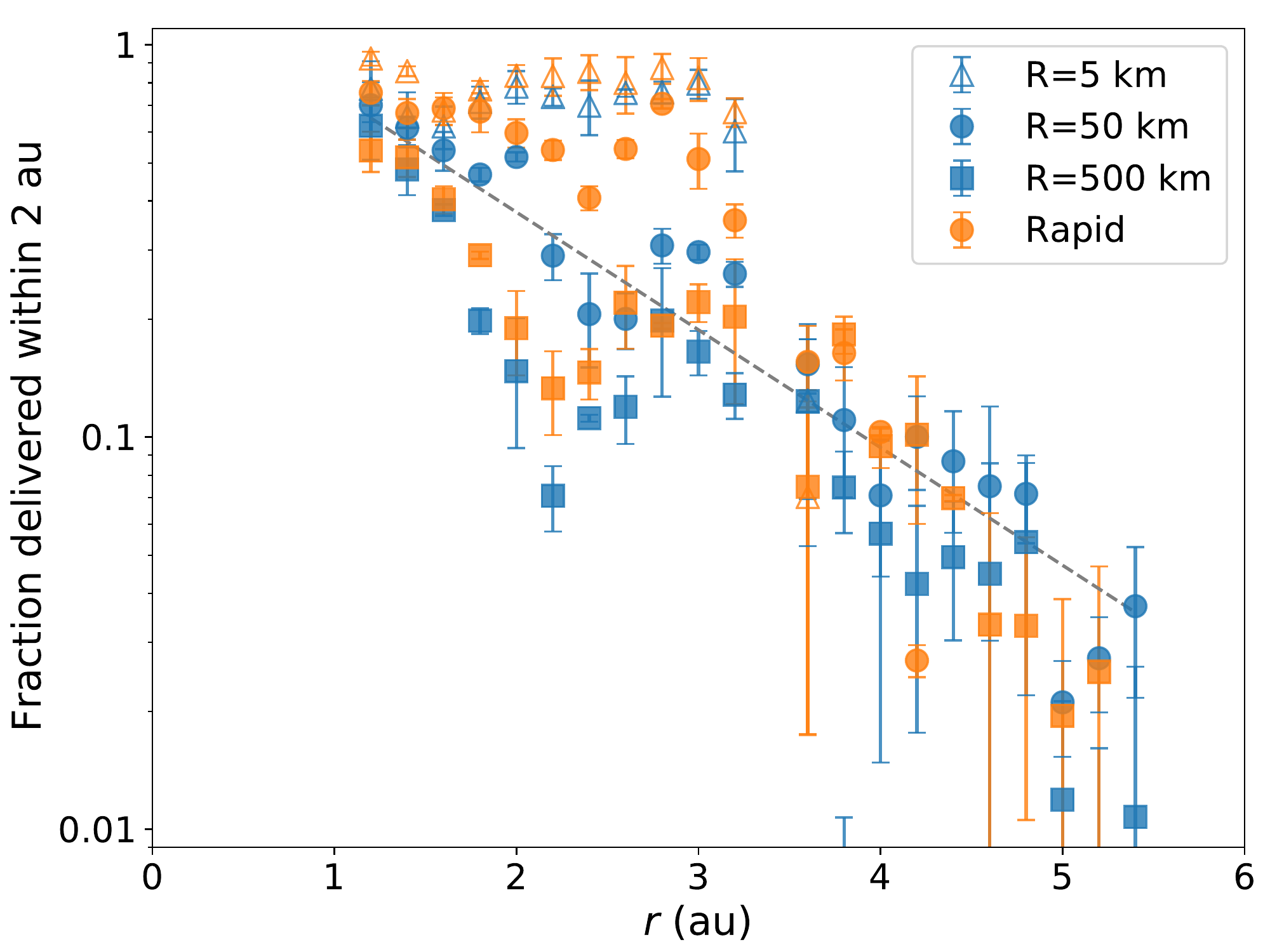}
\end{center}
\caption{Fraction of planetesimals that were originally in each orbit and delivered into the region of $a<2\,{\rm au}$ at the end of simulation ($t=2\,{\rm Myr}$). Blue symbols indicate runs for fiducial Jupiter growth, while orange symbols indicate Jupiter's rapid growth. Error bars represent standard deviations. The dashed gray line shows a fitted equation to all results except for $R=5\,{\rm km}$.}
\label{fig:a_f}
\end{figure}

Furthermore, we considered in more detail the transport from each orbital region. Figure~\ref{fig:a_f} shows the material fraction that was originally in each orbit and transported from each region to the terrestrial planet region at the end of the migration phase ($t=2\,{\rm Myr}$). Except for the case with strong eccentricity damping for $R=5\,{\rm km}$, the transport fraction had a similar trend.
For cases with $R=50\,{\rm km}$ and $500\,{\rm km}$, the transport fraction from each region $f(r)$ can be approximately fitted by the following equation.
\begin{equation}
\label{eq:fraction}
f(r) = (1.49\pm{0.11}) \times 10^{(-0.30\pm{0.018}) r},
\end{equation}
where the plus/minus sign represents the standard error in fitting.
This equation captures the trend relatively well for different sizes and masses of planetesimals except for $R=5\,{\rm km}$. On the other hand, it should be noted that this fitting formula is not universal, as seen in Appendix~\ref{sec:saturn2}, where the slope depends on the orbital configuration of giant planets.

By combining Eq.~(\ref{eq:fraction}) with the initial distribution of materials, the transport for different initial distributions can be calculated.
By assuming the following initial distribution of water,
\begin{eqnarray}
\label{eq:water_distrinbution}
\Sigma_{\rm d,w} = 10 \eta_{\rm w} \left(\frac{r}{{\rm 1\,au}}\right)^{-\frac{3}{2}} {\rm \,g/cm^2},
\end{eqnarray}
where $\eta_{\rm w}$ is the initial water mass fraction of each planetesimal described in Section~\ref{sec:methods}, the amount of water delivery can be calculated by $\int_{1.3\,{\rm au}}^{6\,{\rm au}} f(r) 2 \pi r \Sigma_{\rm d,w} dr$. This yields 31\,$M_{\mathrm{EO}}$ and is consistent with the amount of water delivery shown in Figure~\ref{fig:r_m_mig}.
The calculated values in Figure~\ref{fig:a_f} are obtained under the assumption of initial water distribution. The fitting formula in Eq.~(\ref{eq:fraction}) can be used to estimate the mass transport in any material distribution.
Although water delivery has been the primary focus of this paper so far, the amount of different materials delivered into Earth's orbit can be calculated by providing the material distribution of other elements (e.g., C, N, Mo, D/H ratio).

\section{Discussion}\label{sec:discussion}
\subsection{Early water delivery to the Earth's core}\label{sec:core_water}
This study's results have important implications for the presence of light elements in the Earth's core. The density of the Earth's outer core is 5\%--10\% less than that of pure iron, and is thought to contain large amounts of light elements \citep{1952JGR....57..227B}, such as S, Si, O, C, and H. If a large amount of water were delivered to protoplanets and/or growing Earth, as demonstrated in our numerical simulations, hydrogen could be the most possible candidate for the Earth's core light element. This is because the accretion of planetesimals onto protoplanets and giant impacts among protoplanets induced the formation of magma oceans and/or ponds \citep[e.g.,][]{1993JGR....98.5319T}. Water can easily dissolve into a silicate melt, and hydrogen in the water molecule is largely partitioned into metallic iron during core formation \citep{1997Sci...278.1781O,2021NatCo..12.2588T}.

Hydrogen has some advantages as the Earth's core light elements. For example, a hydrogen-rich outer core is compatible with seismological observations for the density deficit and sound velocity excess \citep{2020E&PSL.53116009U} and compatible with a lower temperature at the core-mantle boundary ($<$ 3,600K) than previously thought ($\sim$ 4,000K) \citep{nomura2014low}.

\begin{figure}[th]
\begin{center}
\includegraphics[width=0.49 \textwidth]{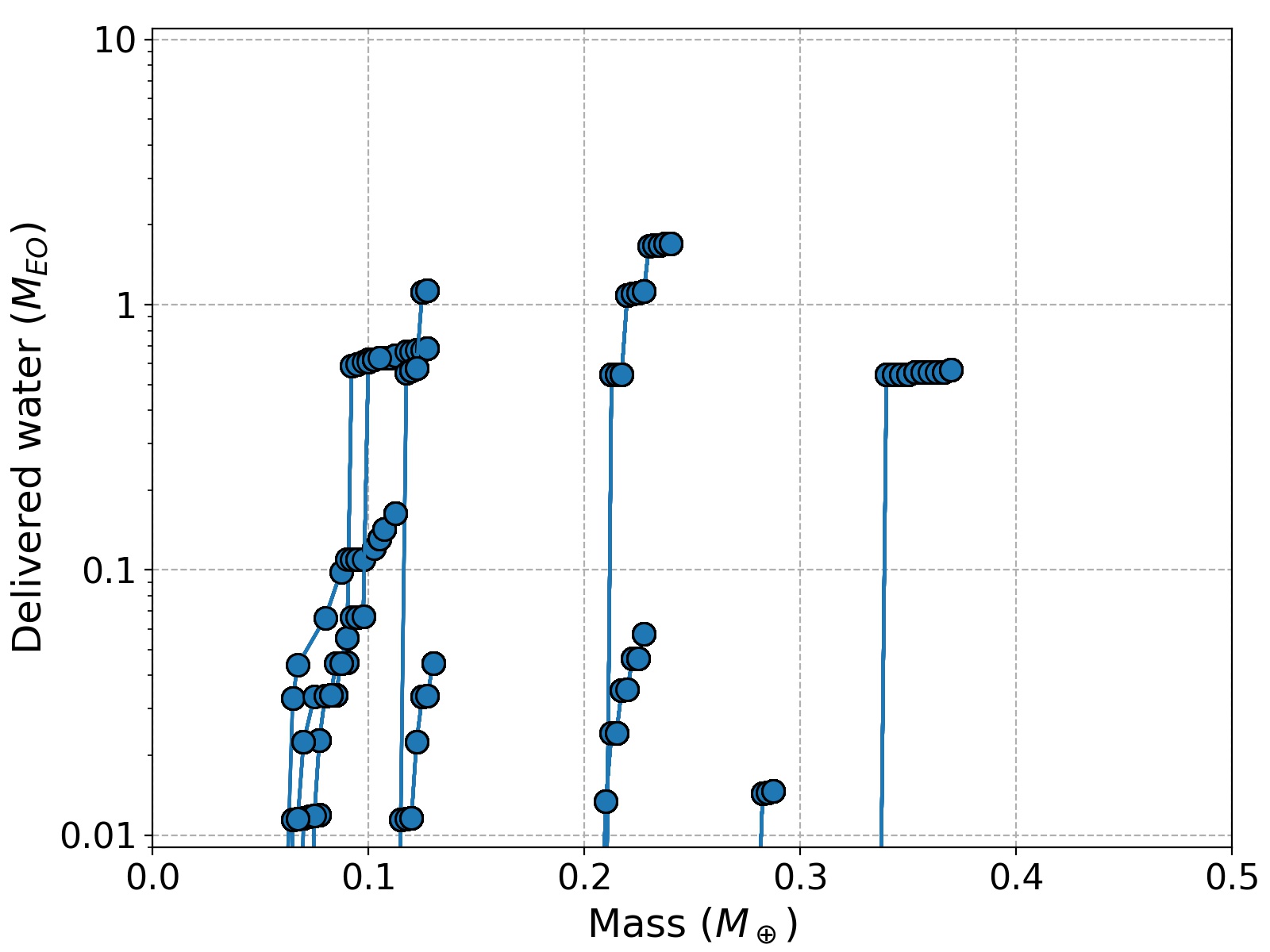}
\end{center}
\caption{Cumulative amount of water delivered to protoplanets as a function of mass of protoplanets for our typical run until $t=2{\rm \,Myr}$. Identical protoplanets are connected by lines. The initial water mass fraction of the planetesimal is the value described in Section~\ref{sec:methods}. Protoplanets initially contains no water.}
\label{fig:delivery}
\end{figure}

If water is delivered to Earth after the last giant impact and/or completion of core formation, which is called the late accretion, there is no chance that hydrogen in water will dissolve into the Earth's core. Figure~\ref{fig:delivery} shows the amount of water transported to protoplanets during the growth and migration stages in our typical simulation for $R=50 {\rm \,km}$ until $t=2{\rm \,Myr}$. Water of several Earth's ocean was found to be transported during the growth of proto-Earth. We continue some simulations up to $t = 10 {\rm \,Myr}$ and see that water-bearing planetesimals accrete on embryos in the relatively early phase of planetary accretion ($t < 10 {\rm \,Myr}$) (see Appendix~\ref{sec:long-term}). During the subsequent giant impact, a magma ocean forms and water dissolves in the mantle, which is later transported to the core\footnote{This is one possible pathway, and there are a range of possible degrees of core-mantle equilibration. However, all siderophile elements (e.g., Mn, Mo, Pt) are depleted in the current Earth's mantle, implying that these elements were extensively transported to the Earth's core during core formation. Therefore, it is likely that if water delivery occurred in the early stage before and during the core formation, hydrogen in water can also be transported to the core.}. Therefore, it is possible that a large amount of water is transported during the early stage of Earth formation, and this may have supplied light elements to the Earth's core.

This study's results suggest a pathway for transporting water to the core. Simultaneously, it is still possible that the ocean that exists today was brought about by late accretion. Notably, it would also be possible that the ocean that exists today was transported during the early stages of Earth's formation, as previously proposed \citep[e.g.,][]{2005Natur.433..842G,2009Natur.461.1227A,2018LPICo2067.6345G}. The transport efficiency of hydrogen in water from the mantle to the core is about 90\%--95\% \citep{2021NatCo..12.2588T} if the pressure at the bottom of the magma ocean is greater than 5\,GPa \citep{1997Sci...278.1781O}, which corresponds to about 500\,km depth for Mars sized planet and about 150\,km depth for Earth sized planet. Therefore, it is possible that the rest of the water became ocean on the surface. According to our simulations, the amount of water delivered to the orbits of terrestrial planets is about 30\,$M_{\mathrm{EO}}$. Longer simulations are needed to discuss how much of this is actually accreted to Earth. If more than one-third of this is accreted to Earth (in fact, according to the results of some simulations continued up to 10\,Myr (Appendix~\ref{sec:long-term}), about $10\,M_{\rm EO}$ are accreted on embryos by $t=10 {\rm \,Myr}$), and about 90\%--95\% of accreted water is transported to the Earth's core and the rest to the mantle and the surface, we would explain both the light elements in the core and the ocean on the surface. In addition, as discussed below in Section~\ref{sec:saturn}, this amount of water can be increased by a factor of about three depending on the initial giant planet configuration. In such a case, a more sufficient amount can be transported. Therefore, even without water delivery by late accretion, water transport during formation may be sufficient to explain the origin of water on Earth. The major component for the late accretion is enstatite-chondrite-like materials \citep{2017Natur.541..521D}, thus transporting carbonaceous-chondrite-like materials by the late accretion would be challenging, which also supports this idea.

We note that the above discussion can be used to estimate the fraction of CC material that constitutes Earth. Assuming that all water is transported by CC planetesimals with a water mass fraction of 5\%, the amount of CC material accreted on Earth is estimated to be $0.14 f\,M_\oplus$, where $f$ indicates the fraction of delivered planetesimals into the terrestrial planet region that actually accrete on Earth. Therefore, Earth accretes about $7f$\% of its mass from CC materials. When the fraction $f$ is assumed to be 0.3--0.5, the fraction of CC material on Earth is estimated about 2\%--4\%, which is approximately consistent with the isotopic study of about 4\% \citep{2021NatAs...5..181B}.

Our results also have implications for water delivery to Venus. Based on hydrogen isotope ratios, Venus has probably contained once as much water as Earth. Following our simulations, the same or slightly less amount of water delivered to Earth would be delivered to protoplanets in Earth's inner orbit. If the pressure at the bottom of the magma ocean was also comparable to that of Earth, a large amount of water might have been transported to the core of Venus, and the remainder might have been present at the surface. A more detailed study of this is needed.

\subsection{Mixing role of Jupiter on NC-CC dichotomy}\label{sec:planet_formation}
Meteorites give us some constraints on Solar System formation. There is a clear dichotomy in the isotopic compositions between the NC and CC groups \citep{2011E&PSL.311...93W}. This is most likely due to the spatial decoupling of the planetesimal formation regions within about 1 Myr after the Solar System birth, and the NC and CC reservoir were separated for at least about 2 Myr until the NC group formation was completed \citep{2017PNAS..114.6712K}. The decoupling here means that the reservoir for forming planetesimals was divided and the transport of planetesimals seems to have been prevented. For example, the influx of CC materials into the inner region is estimated to occur after 4 Myr \citep{2017GeCoA.212..156S}. A meteorite formed by a collision of NC and CC materials has also been found, and the timing of the collision is late and estimated to be approximately 7 Myr \citep{2022M&PS...57..261S}. Notably, the time of $t=0$ (start time of Jupiter's growth) in our \textit{N}-body simulation did not coincide with the timing of Ca-Al-rich inclusions (CAI) formation.

The most widely accepted explanation for this decoupling is that Jupiter acted as a barrier. When Jupiter grows to the pebble isolation mass ($\simeq 20\,M_\oplus$), it stops the radial drift of pebbles \citep{2014A&A...572A..35L}. When Jupiter grows to about $50\,M_\oplus$, it further opens a density gap in the disk gas. Therefore, it is believed that Jupiter's core grew to about $20\,M_\oplus$ by 1 Myr after CAI formation and prevented the NC-CC mixing of pebbles and dusts after that time \citep{2017PNAS..114.6712K}.

\begin{figure}[th]
\plotone{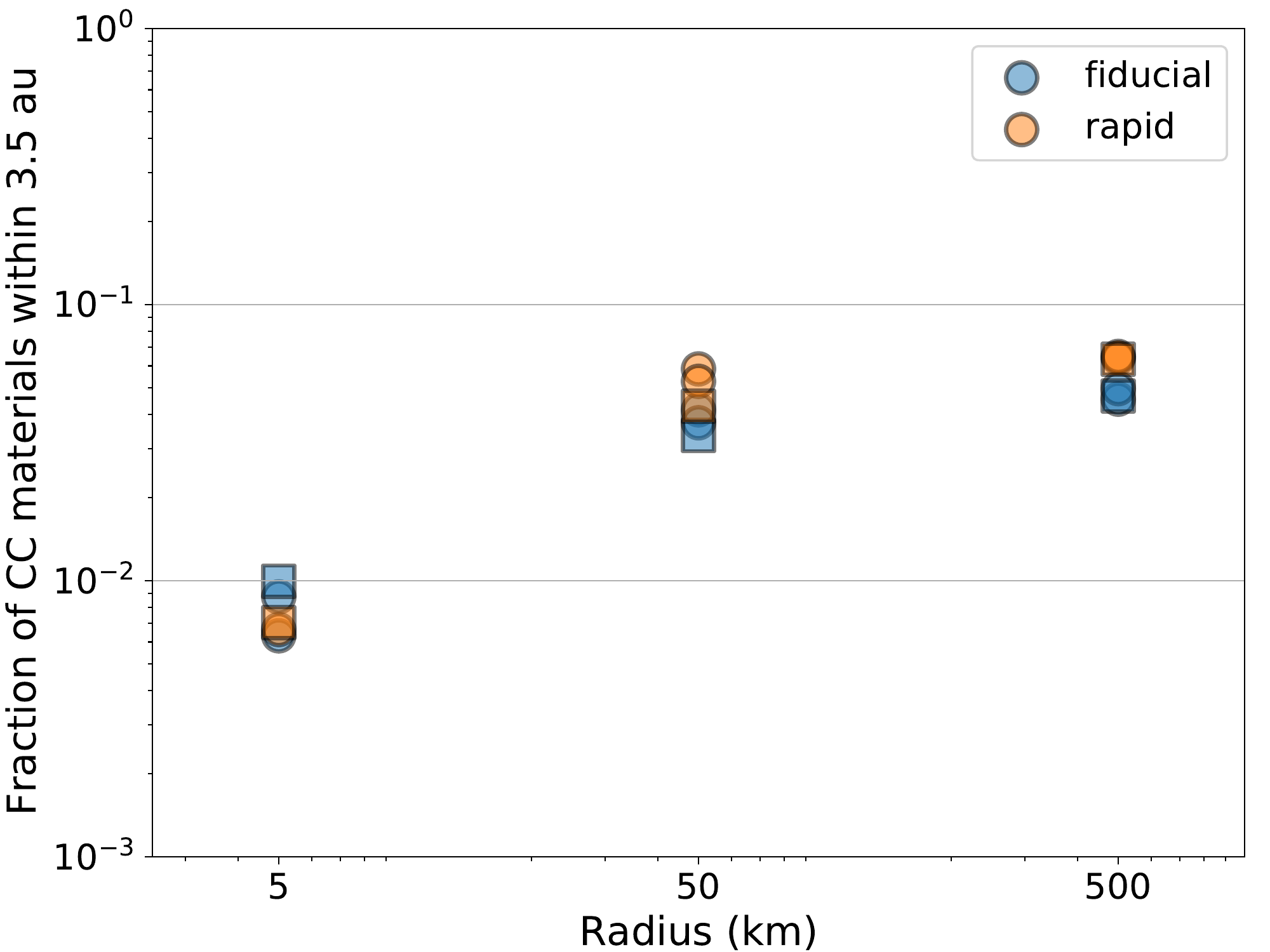}
\caption{The mass fraction of material originating from the outer region (CC materials) among the planetesimals within Jupiter's orbit ($a=3.5\,{\rm au}$) at the end of Jupiter's growth phase ($t=1\,{\rm Myr}$). The materials initially inside and outside of Jupiter's orbit are called NC and CC materials, respectively. Squares represent results for higher resolution simulations.
}
\label{fig:r_CC}
\end{figure}

This study's results show that Jupiter not only acts as a barrier but also mixes with the surrounding material during its growth. Figure\,\ref{fig:r_CC} shows the mass fraction of material transported from the outer region among the planetesimals present in Jupiter's inner orbit at the end of Jupiter's growth phase (before the migration phase). This indicates that about 1-10\% of the material in the inner region comes in from the outer region. This result would not immediately indicate that the NC-CC dichotomy is destroyed by the growth of Jupiter since this mixing fraction is not very high (about 10-20\%, even when Saturn is added to the calculation in Appendix~\ref{sec:saturn2}). Rather, this degree of mixing may be useful in explaining the variation in isotopic composition within each isotopic group \citep{2018Natur.555..507S}.

\subsection{Influence of nearby growing Saturn}\label{sec:saturn}

In this paper, we considered only Jupiter to focus on the effects of Jupiter's growth on material transport. If Saturn grows just outside Jupiter, it would further increase the material mixing. Here, we briefly discussed whether our conclusions would change if we considered the effect of Saturn growing nearby.

We additionally performed simulations with the Saturn core with $M=10\,M_\oplus$ in the orbit $a = 4.5 {\rm \,au}$ as the initial condition. See Appendix~\ref{sec:saturn2} for the simulation results. The additional simulation results confirm that our conclusions remain unchanged (see also Figure~\ref{fig:snap_saturn}). That is, the conclusion that a significant amount of material delivery occurs during the growth phase of giant planets does not change. Notably, the amount of material mixing tended to increase because additional inward scattering occurred due to the presence of Saturn. Regarding the scaling law seen in Figure~\ref{fig:a_f}, we found that the outcome was fitted by a similar formula (see also Figure~\ref{fig:a_f_saturn}), which did not change the finding that early water delivery can deliver water to the Earth's core, as discussed in Section~\ref{sec:core_water}. When the growing Saturn is considered, even more water can be transported in the Earth's growing stage (see also Figure~\ref{fig:delivery_saturn}). Moreover, hydrogen equivalent to about 30--70\,$M_{\mathrm{EO}}$ can be stored in the outer core \citep{2021NatCo..12.2588T}, and this amount is well accounted for.

\subsection{Comparison to previous studies}\label{sec:comparison}
As mentioned in Section~\ref{sec:intro}, there are several studies similar to this study. Here we compare our results with some of them. \citet{2017Icar..297..134R}, which studied the mass transport associated with the growth of giant planets, assumed that Jupiter grows at $a=5.4 {\rm \,au}$ or further, and mainly considered the transport of planetesimals into the asteroid belt. Although their computational setup differs from ours in many ways, a rough comparison is possible. They showed that the influence of growing Jupiter at $a=5.4 {\rm \,au}$ and growing Saturn at $a=7.3 {\rm \,au}$ results in 10--20\% of planetesimals initially in the giant planet region ($a=4-9 {\rm \,au}$) being delivered to the main asteroid belt with perihelion distance below 3.2\,au  (their Fig.~5).
Their results also suggested that the growth of giant planets transports a certain fraction of planetesimals that originally existed in the giant planet region to the terrestrial planet region. From figures in the paper, we can estimate that the fraction of planetesimals delivered to the terrestrial planet region inside $a=2 {\rm \,au}$ is on the order of 1\%.

We compare this to our results of growth phase simulation (Section\,\ref{sec:growth}). According to our standard results, the fraction of planetesimals initially at $a=2-4 {\rm \,au}$ that are delivered into the terrestrial planet region ($a<2 {\rm \,au}$) during the growth of Jupiter is about 10\%. This is about an order of magnitude larger transport than the results in \citet{2017Icar..297..134R}. This difference is interpreted in our simulations as a result of the shorter distance between Jupiter's location ($a=3.5 {\rm \,au}$) and the terrestrial planet region. In fact, the $\sim 10\%$ of the total planetesimals transported in our simulation is similar to the result of \citet{2017Icar..297..134R}, in which about 10--20\% of the planetesimals beyond $a=4 {\rm \,au}$ are transported to the main belt (perihelion distance below 3.2\,au). Therefore, our results are consistent with the results of \citet{2017Icar..297..134R} that mass transport is caused by the growth of giant planets. In addition, our result suggests that orbital location of growing giant planets is important for the water delivery to the terrestrial planet region during the growth phase.

Next, we discuss our results with those of \citet{2014Icar..239...74O}. \citet{2014Icar..239...74O} investigated the long-term orbital evolution of planetesimals with perihelion distances below 2\,au in results of \citet{2011Natur.475..206W}. \citet{2011Natur.475..206W} calculated that a total of $0.0793 \,M_\oplus$ of planetesimals from the giant planet region ($a = 3.5 - 8 {\rm \,au}$) and $0.174 \,M_\oplus$ of planetesimals from orbits outside the giant planets ($a = 8 - 13 {\rm \,au}$) delivered to the terrestrial planet region as a result of the Grand Tack migration. \citet{2014Icar..239...74O} used this as their initial condition. Assuming that the water fraction of the planetesimals is 5\% as in our study, we can estimate that about $17\,M_{\rm EO}$ and $38\,M_{\rm EO}$ of water were delivered to the terrestrial planet region by the Grand Tack migration, respectively.

In our study, the total amount of water transported to the terrestrial planet region after Jupiter's growth and migration is about $10-30\,M_{\rm EO}$  (Figure~\ref{fig:r_m_mig}). This amount is similar to the amount of water inside 2\,au in the initial state of \citet{2014Icar..239...74O}. Therefore, if we continue to follow the long-term evolution of about 100\,Myr, water of about the Earth's ocean mass is expected to be transported to Earth as late accretion, as in \citet{2014Icar..239...74O}.
However, it should be noted that in our study the planetesimals are placed only up to $a=6 {\rm \,au}$. Moreover, considering the presence of Saturn, our simulation delivers about $50-90\,M_{\rm EO}$ of water in the growth and migration stages (Appendix~\ref{sec:saturn2}). Considering these points, the total amount of water delivered to the terrestrial planet region can be higher than in \citet{2014Icar..239...74O} when the growth stage is considered. One of the points of our study is that some water may have been transported during the early phase of planetary accretion and may be stored in the Earth's interior (Section~\ref{sec:core_water}).
In future studies, it may be necessary to perform both simulations that follow the growth and migration stages of giant planets and those that follow the long-term evolution. In such a study, it would be interesting to focus on the dependence of the initial positions of giant planets as suggested above.

\section{Conclusions}\label{sec:conclusions}

Following the Grand Tack model, we investigated material delivery to the terrestrial planet region during Jupiter's growth and migration phases using \textit{N}-body simulations. The simulation results show that the surrounding planetesimals were scattered and a significant amount of planetesimals were transported to the terrestrial planet region during Jupiter's growth. Furthermore, we observed that additional material transport can occur during Jupiter's migration phase. At the end of the migration and growth phase, about 10--40\,$M_{\mathrm{EO}}$ of water are delivered into the terrestrial planet orbits, which was not previously calculated. Although the physical processes of planetesimal delivery depended on the gas drag strength and the rate of Jupiter's growth, we found that the amount of material delivery at the end of the migration phase was less dependent on these parameters.

This study's results have important implications for geochemistry. It is often considered that Earth's oceans have been formed by late accretion after the Earth formation. Water transport may also explain the density deficit in the Earth's core. However, for the transport of water to the core to occur, it may be necessary for water to have been transported during the Earth's formation stage. Our results indicate that large amounts of water could be transported to the primordial Earth during its growing stage. This could explain the origin of the Earth's oceans.
There are also important implications for the NC-CC dichotomy found in meteorites. Previous studies have focused on Jupiter's role as a barrier to the influx of CC materials. Our results indicate that Jupiter's growth is responsible for mixing some NC and CC materials, which may explain the variation in isotopic composition within the NC and CC groups.

Several assumptions were used in the simulations in this study. We assume that giant planet cores do not migrate during growth. The rate and direction of orbital migration depend on a complex set of conditions \citep[e.g.,][]{2019MNRAS.484..728M}. In a separate study, we will investigate whether our results change when the core moves.
In addition, this study is limited to the region up to $a = 6 {\rm \,au}$. To make further effective use of the results of planetary exploration, as represented by Ryugu, simulations with a wider computational domain are needed. Such simulations will also be the subject of future work.

\begin{acknowledgments}
We thank the two anonymous reviewers for helpful comments.
Numerical computations were performed in part on a PC cluster at the Center for Computational Astrophysics of the National Astronomical Observatory of Japan.
This work was supported by MEXT KAKENHI grant Nos. JP17H06454, JP17H06456, JP17H06457, JP18K13608 and JP19H05087.
\end{acknowledgments}

%

\appendix

\section{Subsequent simulations after the Grand Tack}\label{sec:long-term}
The purpose of this work is to clarify water transport through planetesimals during the growth and migration phases of giant planets. Whether planetesimals delivered to the terrestrial planet region actually accrete on the growing Earth requires longer-term \textit{N}-body simulations \citep[e.g.,][]{2014Icar..239...74O}. Such simulations need to be done in a separate study, but we have continued the simulations in some fiducial case simulations up to $t=10{\rm \,Myr}$.

\begin{figure}[th]
\epsscale{0.5}
\plotone{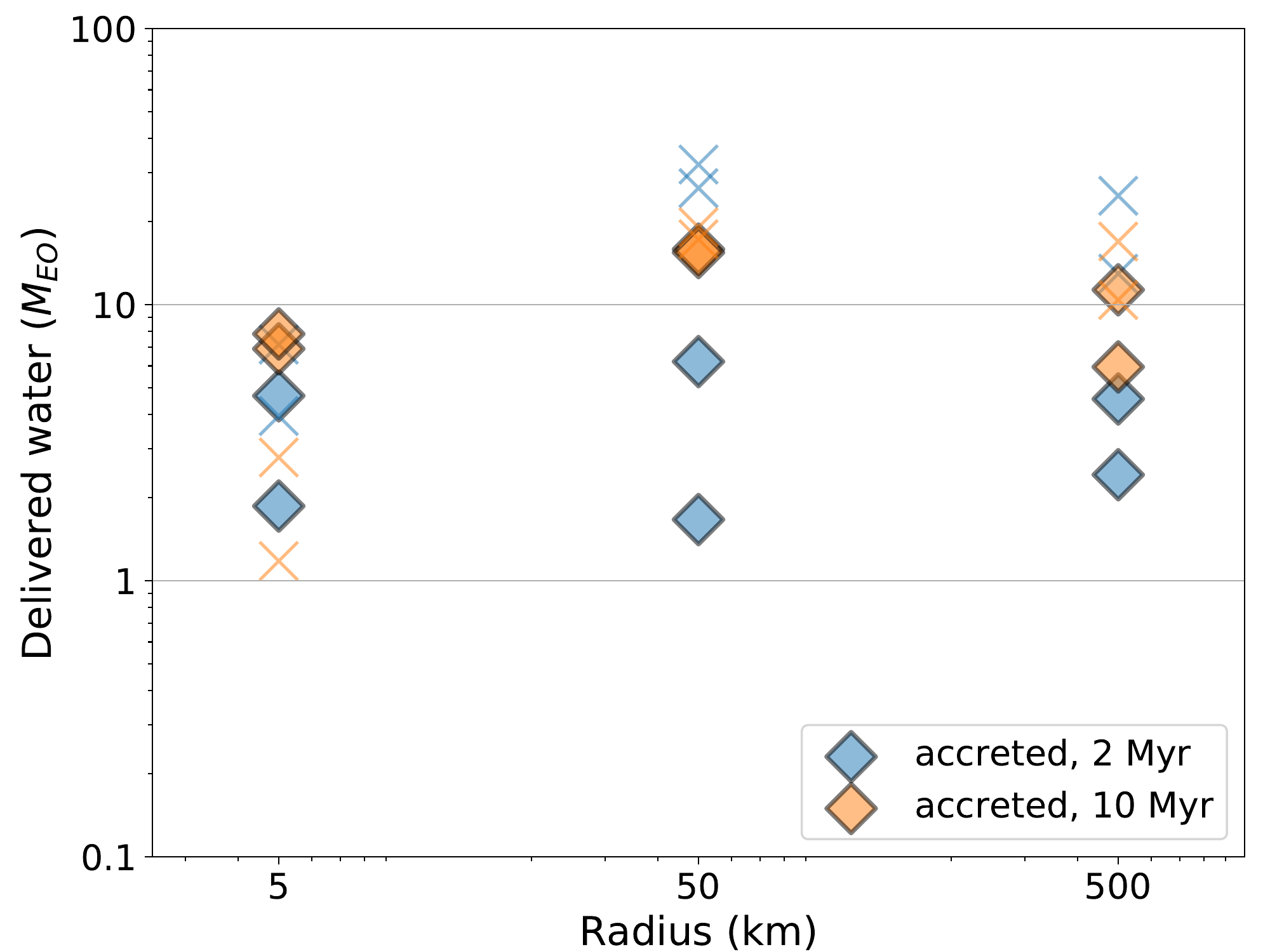}
\caption{The amount of water delivered into the region of $a < 2 {\rm \,au}$ until $t=2{\rm \,Myr}$ (blue) and $t=10{\rm \,Myr}$ (orange) for the fiducial case, corresponding to blue circles in Figure~\ref{fig:r_m_mig}. Diamonds represent the amount of water accreted on embryos, and crosses represent the amount of water present as planetesimals.}
\label{fig:r_m_mig_10Myr}
\end{figure}

Figure~\ref{fig:r_m_mig_10Myr} shows the amount of water present in the terrestrial planet region at $t = 2 {\rm \,Myr}$ and 10\,Myr. At $t = 2 {\rm \,Myr}$, the amount of water accreted on the embryos is less than $6\,M_{\rm EO}$, but at $t = 10 {\rm \,Myr}$, about $10\,M_{\rm EO}$ of water is accreted on terrestrial planets. Although \citet{2014Icar..239...74O} showed that water accretion by terrestrial planets is in the late phase of terrestrial planet accretion (after 10\,Myr after the Grand Tack migration), this result suggests that water accretion on terrestrial planets may be in a relatively early phase of planetary accretion. More detailed investigations should be performed in a future study.

\section{Additional simulations with Saturn}\label{sec:saturn2}

As additional simulations, we performed \textit{N}-body simulations in the case where Saturn grows near Jupiter.
In addition to the Jupiter core at $a=3.5 {\rm \,au}$, we started the simulation with the Saturn core of $M=10\,M_\oplus$ at $a=4.5 {\rm \,au}$. The Saturn core is assumed to take approximately 1 Myr to grow to Saturnian mass. We put 1019 planetesimals with $M=0.005\,M_\oplus$ between $a=1.3 - 6 {\rm \,au}$, excluding the very vicinity of giant planets. The gas drag is calculated assuming $R=50 {\rm \,km}$.
The Grand Tack model is mimicked for the migration of the giant planets. During the outward migration, Jupiter and Saturn are in a 3:2 resonant configuration.

\begin{figure*}[th]
\epsscale{0.7}
\plotone{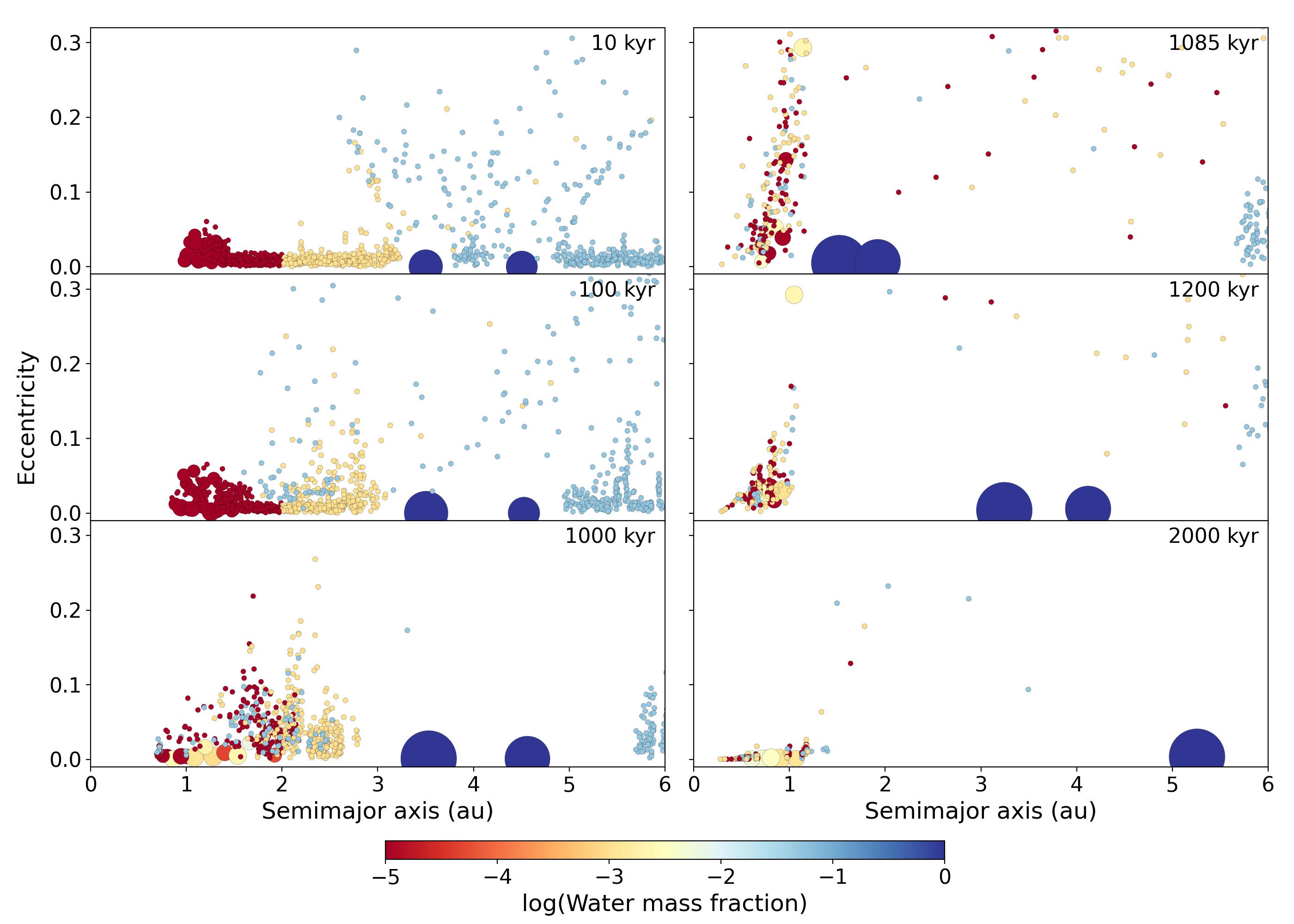}
\caption{Snapshots of the system during the growth and migration phases in a case with Saturn. Jupiter and Saturn grow on a timescale of 1 Myr, and move inward and outward according to the Grand Tack model.}
\label{fig:snap_saturn}
\end{figure*}

Figure~\ref{fig:snap_saturn} represents snapshots in the growth and migration phases. In addition to the scattering of planetesimals associated with Jupiter's growth, the scattering associated with Saturn's growth enhances material transport. At the end of the growth phase, water of about 70-90$\,M_{\mathrm{EO}}$ is transported to the region within $2 {\rm \,au}$. Additional transport occurs during the migration phase of giant planets. At the completion of migration, water of about 50-90$\,M_{\mathrm{EO}}$ is delivered.

\begin{figure}[th]
\epsscale{0.5}
\plotone{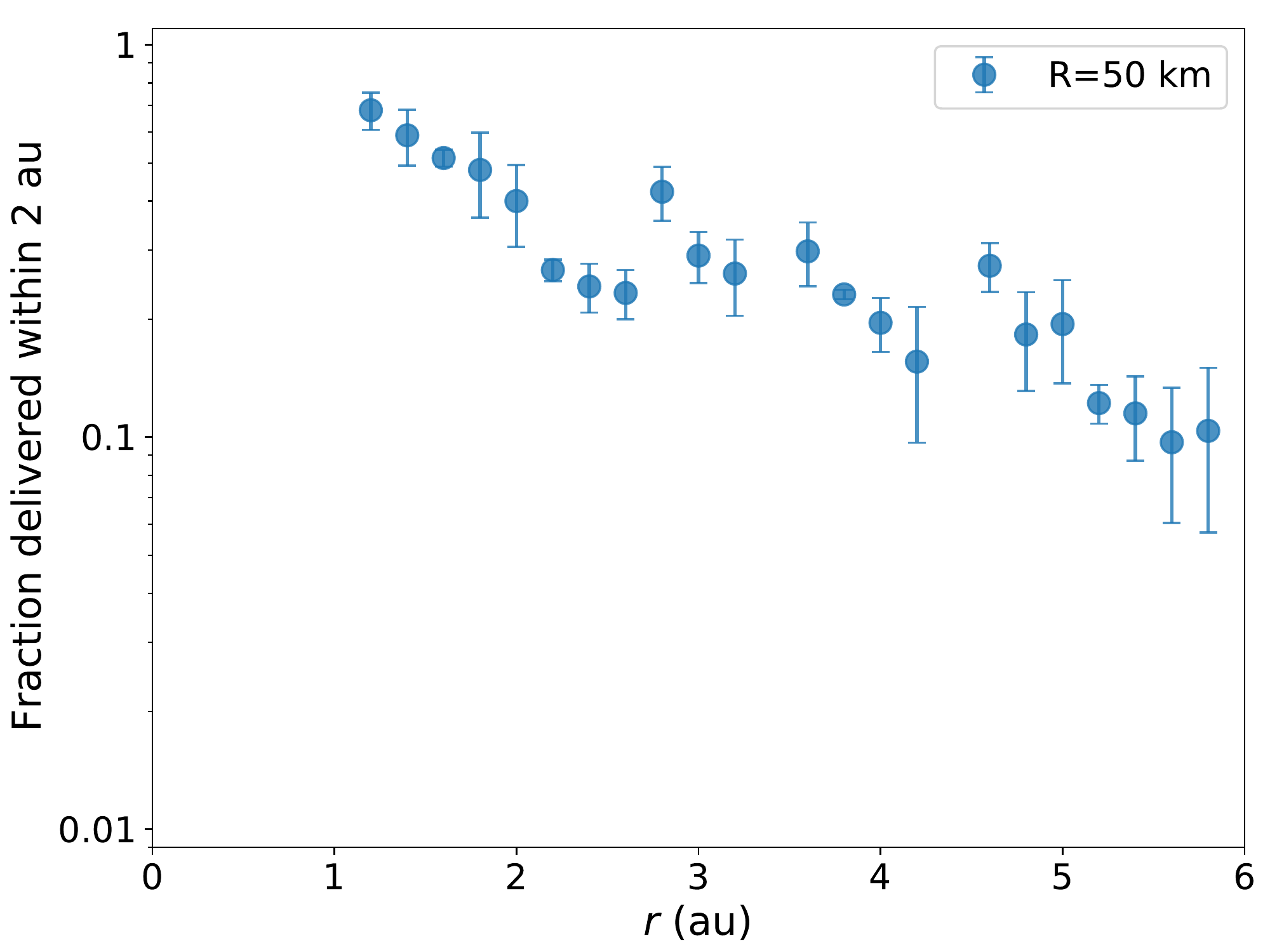}
\caption{Same as Figure~\ref{fig:a_f} but for the case with Saturn. Three runs of simulations are summarized.}
\label{fig:a_f_saturn}
\end{figure}

Figure~\ref{fig:a_f_saturn} represents the fraction of planetesimals initially present in each region that are transported within $a=2 {\rm \,au}$ after the end of simulation. Similar to what is seen in Figure~\ref{fig:a_f}, the outcome can be fitted by a similar function to Eq.~(\ref{eq:fraction}). The overall delivery efficiency is increased, and the slope is slightly shallower.

\begin{figure}[th]
\epsscale{0.5}
\plotone{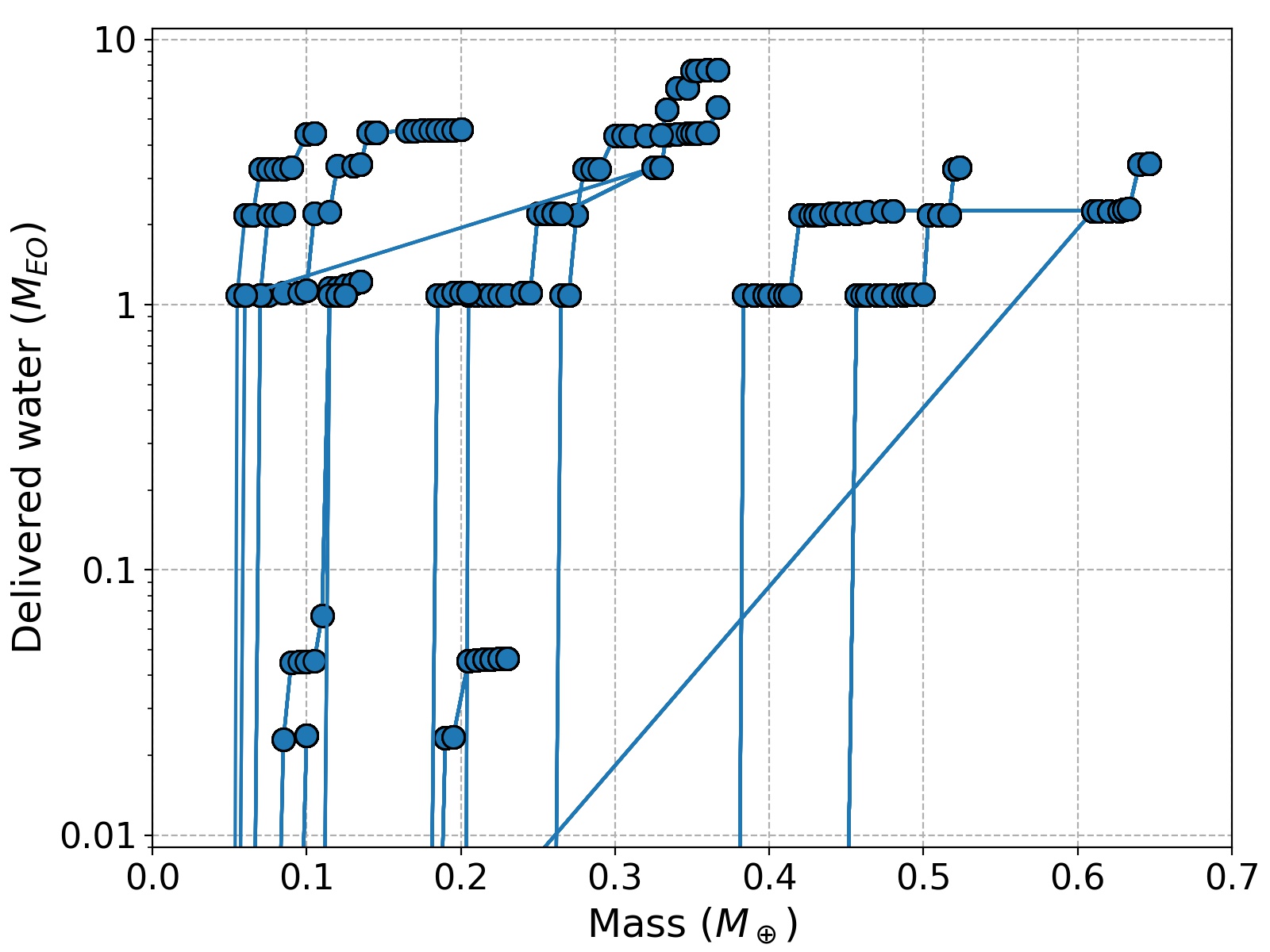}
\caption{Same as Figure~\ref{fig:delivery} but for the case with Saturn.}
\label{fig:delivery_saturn}
\end{figure}

Figure~\ref{fig:delivery_saturn} shows the amount of water transported to each protoplanet for one run. Compared to the Jupiter-only case (Figure~\ref{fig:delivery}), the early water delivery is more pronounced. The water accreted during the early stages of Earth formation can be dissolved into the core to explain the core density deficit, and it may also explain the water on the Earth surface.


\bibliography{mybibfile}{}
\bibliographystyle{aasjournal}



\end{document}